\documentclass{cimento}

\usepackage{graphicx}
\usepackage{amssymb}
\title{Likelihood Analysis of GRB Evolution with Redshift}
\author{
C.~Graziani\from{ins:uofc},
T.~Q.~Donaghy\from{ins:uofc}, \atque
D.~Q.~Lamb\from{ins:uofc}
}
\instlist{\inst{ins:uofc} Department of Astronomy \& Astrophysics,
University of Chicago}
\PACSes{
\PACSit{98.70.Rz}{gamma-ray sources; gamma-ray bursts}
}
\newcommand{\epk}{\ensuremath{E_{pk}}}
\newcommand{\ypk}{\ensuremath{y_{pk}}}
\newcommand{\ya}{\ensuremath{y_{A} }}
\newcommand{\spk}{\ensuremath{\sigma_{pk}}}
\newcommand{\sa}{\ensuremath{\sigma_{A}}}
\newcommand{\ypkli}
{\ensuremath{y_{pk}^{(l,i)}}}
\newcommand{\yali}
{\ensuremath{y_{A}^{(l,i)}}}

\newcommand{\zli}
{\ensuremath{z^{(l,i)}}}
\newcommand{\xa}{\ensuremath{x_{A} }}
\newcommand{\xpk}{\ensuremath{x_{pk} }}

\newcommand{\eiso}{\ensuremath{E_{iso}}}
\newcommand{\xiso}{\ensuremath{x_{iso}}}
\newcommand{\epksrc}
{\ensuremath{E_{pk}^{(src)}}}
\newcommand{\xpksrc}
{\ensuremath{x_{pk}^{(src)}}}
\newcommand{\egamma}{\ensuremath{E_{\gamma}}}
\newcommand{\xgamma}{\ensuremath{x_{\gamma}}}
\newcommand{\barxgamma}{\ensuremath{\bar{x}_{\gamma}}}
\newcommand{\tjet}{\ensuremath{\theta_{jet}}}
\newcommand{\ojet}{\ensuremath{\Omega_{jet}}}
\newcommand{\xjet}{\ensuremath{x_{jet}}}

\newcommand{\sfrfn}{\ensuremath{\mu}}
\newcommand{\xgfn}{\ensuremath{\omega}}
\newcommand{\xjfn}{\ensuremath{\kappa}}
\newcommand{\xpkfn}{\ensuremath{\zeta}}

\begin{document}

\maketitle

\begin{abstract}
We present a likelihood approach to modeling multi-dimensional GRB
\epk--fluence--redshift data that naturally incorporates instrument
detection thresholds.  The treatment of instrument thresholds is essential
for analyzing evidence for GRB evolution.  The method described here
compares the data to a uniform jet model, in which the jet parameters are
allowed to vary with redshift.

Data from different experiments may be modeled jointly.  In
addition, BATSE data (for which no redshift information is available) may
be incorporated by ascribing to each event a likelihood derived from the
full model by integrating the probability density over the unknown
redshift.  The loss of redshift information is mitigated by the large
number of available bursts.

We discuss the implementation of the method, and validation of it using
simulated data.

\end{abstract}

\section{Introduction}

Analyses of HETE-2- and BeppoSAX-localized Gamma-ray bursts with known
redshifts provide evidence that the isotropic-equivalent GRB energy \eiso
evolves with redshift \cite{gr2003}.  The assessment of the strength of
this evidence is linked with the issue of detection thresholds, which can
create apparent energy-redshift correlations by truncation.  The modeling
of such thresholds is a complex and delicate matter

We present here a method, based on the likelihood function, for analyzing
GRB \epk--fluence--redshift data in the presence of instrument thresholds.
We have developed the method with a view to addressing the question of
whether GRB properties do, in fact, evolve with redshift.

\section{The Method}

We compare observations of GRB data with a model of GRB source properties,
using the likelihood function.   

The data consists of GRB fluences and \epk  values, with errors, and GRB
afterglow redshifts, where available.  We use the results of fits of the
Band function to integrated GRB spectra.  The fit parameters that we use as
data are $\ypk\equiv\log_{10}(\epk)$ and $\ya\equiv\log_{10}(A)$, where
$A$ is the amplitude --- the scale factor that multiplies the Band
function.  These parameters are supplemented by their errors, \spk and
\sa, respectively.

The model is a generalization of the variable-angle uniform jet model of
Lamb, Donaghy and Graziani \cite{l2004}.  The generalization is that the
parameters of the distribution of jet opening angles are allowed to evolve
with redshift.

The observables of the model are:

\begin{itemize}

\item $z$, the redshift.

\item $\xiso\equiv\log_{10}(\eiso/10^{52}~\mbox{erg})$.

\item $\xpksrc\equiv\log_{10}(\epksrc/1~\mbox{keV})$, where \epksrc is the
\epk\  parameter in the source frame.

\item $\xjet\equiv\log_{10}(\ojet/4\pi)\equiv\log_{10}(1-\cos\tjet)$,
where $\tjet$ is the opening half-angle of the uniform jet.

\item $\xgamma\equiv\log_{10}(\egamma/10^{52}~\mbox{erg})$, where \egamma\
is the prompt gamma-ray energy actually emitted:  $\xgamma=\xiso+\xjet$. 

\end{itemize}

The model is introduced through the differential event rate:
\begin{eqnarray}
\rho(z,\xgamma,\xjet,\xpksrc)dz\,d\xgamma\,&d\xjet&\,d\xpksrc =\nonumber\\
&&\sfrfn(z)dz \times \xgfn(\xgamma)d\xgamma \times
\xjfn(\xjet|z)d\xjet \times \xpkfn(\xpksrc|\xiso)d\xpksrc.
\label{diff}
\end{eqnarray}
where

\begin{itemize}

\item $\sfrfn(z)$ is proportional to the
Rowan-Robinson \cite{rr2001} SFR function, multiplied by $(1+z)^{-1}
dV(z)/dz$, where $V(z)$ is the proper volume of the sphere bounded by
redshift $z$.

\item
$\xgfn(\xgamma)\equiv
\exp\left(-\frac{1}{2}\frac{(\xgamma-\barxgamma)^2}{\Gamma^2}\right)$, with
$\Gamma = 0.33$ \cite{l2004} and \barxgamma\ a fit parameter.

\item
$\xjfn(\xjet|z)\equiv{\ojet}^{-\delta}\times\frac{d\ojet}{d\xjet}$,
$\ojet^{(1)}<\ojet<\ojet^{(2)}$.  $\delta$, $\ojet^{(1)}$, and
$\ojet^{(2)}$ may be functions of redshift.

\item
$\xpkfn(\xpksrc|\xiso)\equiv
\exp\left(-\frac{1}{2}\frac{(\xpksrc-\frac{1}{2}\xiso-b)^2}{\Lambda^2}\right)$,
with $b = 1.95$ and $\Lambda = 0.13$, is the broadened Amati relation
\cite{amati2002,l2004}.

\end{itemize}

The likelihood function is given by
\begin{equation}
L=\prod_{l=1}^{N_{\mbox{\scriptsize instruments}}}\prod_{i=1}^{N_l}
\left\{\frac{f(\yali,\ypkli,\zli)}{\bar{N_l}}\right\},
\label{lfn}
\end{equation}
where $\bar{N_l}$ is the expected number of
events observed by the $l$-th instrument, given its specific threshold,
and $f(\yali,\ypkli,\zli)$ is
the individual likelihood of the $i$-th event observed by the $l$-th
instrument.

In Eq. (\ref{lfn}), the expected number of events $\bar{N}$ is given by
\begin{equation}
\bar{N_l}=\int
dz\,d\xjet\,d\xgamma\,d\xpksrc\,\rho(z,\xgamma,\xjet,\xpksrc)\times
(\ojet/4\pi)\times\eta_l(z,\xiso,\xpksrc).
\end{equation}
Note that: 
\begin{itemize}

\item $\eta_l(z,\xiso,\xpksrc)$ is the detection
probability by the $l$-th instrument of a GRB whose vital statistics are
$z, \xiso, \xpksrc$.  It is here that the instrumental threshold is
incorporated.  $\eta$ is given by an analytic function that depends on the
instrument thresholds.  We adopt the threshold values of Band \cite{b2003}.

\item
The factor $\ojet/4\pi$ is the probability that our line-of-sight lies inside
the jet.

\item
$\bar{N_l}$ is reducible to a three-dimensional integral over $z$, \xiso,
and \xpksrc. This integral must be performed numerically.

\end{itemize}

In Eq. (\ref{lfn}), the event likelihood $f$ is expressed in terms of
variables  $\xa\equiv\log_{10}(A)$,
$\xpk\equiv\log_{10}(\epk)$, which are related to \xiso, \xpksrc by
\begin{eqnarray}
\xpksrc&=&\xpk+\log_{10}(1+z)\nonumber\\
\xiso&=&\xa+(\alpha+2)\xpk+\log_{10}\left(\frac{d_L(z)^2}{1+z}\right)-\lambda\nonumber,
\end{eqnarray}
where $\alpha$ is the Band-function low-energy power-law index (assumed
equal to -1 throughout), $\lambda$ is a unit-conversion constant, and
$d_L(z)$ is the luminosity distance.  We then have
\begin{eqnarray}
f(\ya,\ypk,z)=&\int&d\xjet\,d\xpk\,d\xa\,
               \rho(z,\xgamma=\xiso+\xjet,\xjet,\xpksrc)\nonumber\\
&&\times\exp\left[-\frac{1}{2}\frac{(\ya-\xa)^2}{\sa^2}
            -\frac{1}{2}\frac{(\ypk-\xpk)^2}{\spk^2}\right].\nonumber
\end{eqnarray}
This integral may be performed analytically by standard techniques.

In the case of events for which no redshift is known, the event
likelihood should be replaced in the full likelihood by its marginalized
version, i.e. $f(\ya,\ypk,z)\rightarrow \int_0^\infty dz\,f(\ya,\ypk,z)$.

\section{Implementation and Validation}

GRB evolution is currently implemented by allowing the three parameters of
$\xjfn(\xjet|z)$ (power-law index, lower- and upper-bounds of power-law
domain) to be themselves power-laws in $1+z$.  Evolution may be turned off
by setting the indices of these power-laws to zero.

The resulting six parameters are supplemented by a seventh ---
\barxgamma. 

The likelihood function is interpreted according to the principles of
Bayesian inference.  It is converted to a posterior density by
multiplication by an assumed prior density on the model parameters. The
shape of the resulting probability distribution in parameter space may be
used to estimate (or set constraints upon) the parameter values. The peak
of the distribution constitutes the best-fit parameter estimate, while
confidence regions may be obtained by considering level curves of the
posterior density containing prescribed amounts of probability (90\%,
say).

\begin{figure}[t]
\centerline{
\includegraphics[scale=0.4]{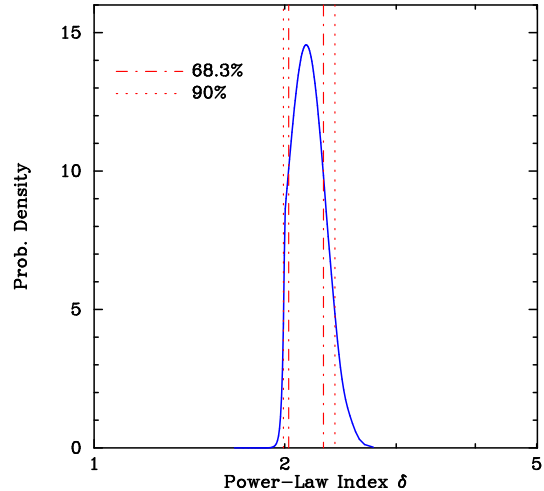}
}
\caption{Estimating the power-law index of the $\ojet$ distribution.}
\label{f1}
\end{figure}

We are currently validating the code by running it on samples of simulated
(non-evolving) GRBs produced by the simulator described in \cite{l2004}.
Figure \ref{f1} shows the result of estimating the power-law index of the
$\ojet$ distribution using 25 events drawn from a simulation that assumes
an index $\delta=2$.  The true value of the index lies within the 90\%
confidence interval.

Work on estimating \barxgamma\ and the other
power-law parameters (which are more broadly distributed and more highly
correlated) is in progress.  We will be applying the experience obtained
from analyzing simulations to real data from HETE, BeppoSAX and BATSE in
the near future.

\end{document}